\documentclass[referee]{raa}            

\usepackage{graphicx,times}             
\usepackage{natbib}
\usepackage{amssymb,amsmath}
\bibpunct{(}{)}{;}{a}{}{,}

\usepackage[pagebackref=true]{hyperref}

\begin{document}

  \title{Observational constraints on phenomenological emergent dark energy and barotropic dark matter characterized by a constant equation of state parameter}

   \volnopage{Vol.0 (20xx) No.0, 000--000}      
   \setcounter{page}{1}          

   \author{Jian-Qi Liu 
      \inst{}
   \and Yan-Hong Yao
      \inst{}
   \and Yan Su
      \inst{}
   \and Jia-Wei Wu
   \inst{}
   }

   \institute{School of Physics and Astronomy, Sun Yat-sen University, 2 Daxue Road, Tangjia, Zhuhai, 519082, People's Republic of China; {\it yaoyh29@mail.sysu.edu.cn}\\
\vs\no
   {\small Received 20xx month day; accepted 20xx month day}}

\abstract{ While cold dark matter is widely supported by a range of cosmological observations, it encounters several difficulties at smaller scales. These issues have prompted the investigation of various alternative dark matter candidates, leaving the question "What is dark matter?" still open. In this work, we propose a new cosmological model that considers dark matter as a barotropic fluid with a constant equation of state parameter and interprets dark energy as the phenomenological emergent dark energy rather than a cosmological constant. We then place constraints on our new model using the Planck 2018 Cosmic Microwave Background (CMB) anisotropy measurements, Baryon Acoustic Oscillation (BAO) measurements from the Dark Energy Spectroscopic Instrument (DESI), the Pantheon Plus (PP) compilation of Type Ia supernovae (Ia SNe), and the Redshift Space Distortions (RSD) data from Gold2018. The results show   statistically significant signal for  positive dark matter equation of state and square of sound speed  $w_{\rm dm}=c_{\rm s,dm}^2$ ($10^{7}w_{\rm dm}$ = $4.0^{+2.5}_{-2.3}$ at the 95\% confidence level) for the data combination CMB+DESI+PP+RSD. However, Bayesian evidence indicates that this data combination favors the $\Lambda$CDM model with very strong evidence. 
}

   \authorrunning{J.-Q. Liu et al.}            
   \titlerunning{New cosmological model with barotropic dark matter}  

   \maketitle

%
%
\section{Introduction}           
\label{intro}
Dark matter (DM), a mysterious form of matter that does not interact with photons, is estimated to account for around one-quarter of the total energy in the universe. Since the nature of DM remains uncertain, contemporary scientists tend to study DM in a phenomenological way. Among them, the most widely studied phenomenological DM model is the cold dark matter (CDM) model, in which CDM is assumed to behave like a non-interacting ideal fluid, possessing a zero equation of state (EoS) parameter, sound speed, and viscosity. Accompanied by a cosmological constant, CDM is supported by a wide range of cosmological data~\citep{riess1998observational,perlmutter1999measurements,dunkley2011atacama,hinshaw2013nine,story2015measurement,alam2017clustering,troxel2018dark,aghanim2020planck-1}. However, this paradigm encounters challenges at small scales, including issues such as the missing satellite problem~\citep{klypin1999missing, moore1999dark}, the too-big-to-fail dilemma~\citep{boylan2012milky}, and the core-cusp issue~\citep{moore1999cold, springel2008aquarius}. These small-scale problems have led to the consideration of alternative DM models, including warm DM~\cite{blumenthal1982galaxy, bode2001halo}, fuzzy DM~\cite{hu2000fuzzy, marsh2014model}, interacting DM~\cite{spergel2000observational}, and decaying DM~\cite{wang2014cosmological}. These new candidates aim to reduce the formation of low-mass structures while remaining consistent with large-scale observational data.

All DM candidates can be studied in a phenomenological way within the framework of generalized dark matter (GDM). The GDM first proposed in~\cite{hu1998structure} (for follow-up studies on the GDM model, we recommend taking a look at ~\cite{mueller2005cosmological, kumar2014observational, armendariz2014cold, kopp2018dark, kumar2019testing, ilic2021dark, pan2023iwdm}) and is specified by the DM EoS, the rest-frame sound speed, and the viscosity. Since there have already been many work on the extended properties of DM in the context of a cosmological constant~\cite{mueller2005cosmological,kopp2018dark,kumar2019testing,ilic2021dark,armendariz2014cold}, in this work we will follow the approach of Ref.~\citep{yao2024observational}, i.e., considering the phenomenological emergent dark energy (PEDE) as the invisible sector that accelerates the expansion rate of the current universe, considering some interesting results presented at the end of this paper. The PEDE model, initially proposed by the authors of Ref.\cite{li2019simple} to address the Hubble tension~\cite{DiValentino:2020zio,DiValentino:2021izs,Schoneberg:2021qvd,Shah:2021onj,Cai:2021weh,Cai:2022dkh,Huang:2024erq,Huang:2024gfw,Liu:2019awo,Huang:2021tvo,Luo:2020ufj}, has been subsequently extended to a generalized parameterization form capable of accommodating both the cosmological constant and the PEDE model\cite{li2020evidence} (also see~\cite{yang2021generalized,hernandez2020generalized}). Notably, the PEDE model features an identical number of free parameters as the spatially flat $\Lambda$CDM model. This framework was further examined in Ref.~\cite{pan2020reconciling}, which explored its complete evolution encompassing both background and perturbations, yielding results consistent with previous works regarding the $H_0$ tension by using recent observational datasets. In the Ref.~\citep{yao2024observational}, the authors set the DM EoS parameter as a free parameter and the DM sound speed to zero in the context of PEDE. After placing constraints on this model with some combinations of the Planck 2018 Cosmic Microwave Background (CMB) anisotropy measurements, baryon acoustic oscillation (BAO) measurements, and the Pantheon compilation of Type Ia supernovae (Ia SNe), the final results show statistically significant signal for  a negative DM EoS parameter for CMB+Pantheon and CMB+BAO+Pantheon datasets, which is very interesting and worth further investigation. In order to investigate whether introducing other GDM parameters in the context of PEDE would lead to different interesting results, in this paper, building upon the assumption of keeping the DM EoS parameter as a free parameter, we further assume that the DM sound speed is a parameter and hence not being fixed at zero. For the sake of model simplicity, we assume that the DM non-adiabatic sound speed and viscosity are zero, i.e., DM is barotropic. Then we will constrain the new model using Planck2018 CMB anisotropy measurements, BAO measurements from the Dark Energy Specroscopic Instrument (DESI), the Pantheon Plus (PP) compilation of Ia SNe, and the Redshift Space Distortions (RSD) data from Gold2018, and analyze the fitting results of relevant free parameters to see if there are indeed some interesting outcomes emerging.

The structure of this paper is as follows. Section~\ref{methods} provides an overview of the key equations in our new model. Section~\ref{data} introduces the observational datasets and the statistical approach used. In section~\ref{results}, we present the constraints from the observational data and discuss the implications of our model. The final section offers concluding remarks.


\section{Introduction of the PEDE+$\bm{w}$DM Model}\label{methods}
This study assumes a spatially flat, homogeneous, and isotropic spacetime, represented by the spatially flat Friedmann-Robertson-Walker (FRW) metric. We postulate that general relativity appropriately describes the gravitational sector, where matter is minimally coupled to gravity. Additionally, we assume that the fluids in the universe do not interact non-gravitationally and that the universe consists of radiation, baryons, GDM, and PEDE. Consequently, we can express the dimensionless Hubble parameter as
\begin{align}
E^2=\frac{H^{2}}{H_{0}^{2}} =  \hspace{0.1cm} \Omega_{\mathrm{r0}}(1+z)^{4} + \Omega_{\mathrm{dm0}}(1+z)^{3(1+w_{\mathrm{dm}})} + \Omega_{\mathrm{b0}}(1+z)^{3} + \Omega_{\mathrm{de}}(z),
	\label{Eq:H}
\end{align}
where $H$ is the Hubble parameter, $\Omega_{\mathrm{r0}}$, $\Omega_{\mathrm{dm0}}$, $\Omega_{\mathrm{b0}}$, and $\Omega_{\mathrm{de}}$ are the density parameters for radiation, GDM, baryons, and PEDE respectively, here $\Omega_{\mathrm{de}}$ is parameterized in the following form~\cite{li2019simple,pan2020reconciling}:
\begin{align}
	\Omega_{\mathrm{de}}(z) = \Omega_{\mathrm{de0}}[1-\tanh(\log_{10}(1+z))].
	\label{Eq:Omde}
\end{align}
where $\Omega_{\rm de0}=1-\Omega_{\rm r0}-\Omega_{\rm dm0}-\Omega_{\rm b0}$ and $1+z=a^{-1}$(Note that we have normalized the present-day scale factor to 1.) Given our assumption that the fluids do not interact non-gravitationally, we consider them as separate entities; therefore, the PEDE conservation equation reads
\begin{align}
	\dot{\rho_{\mathrm{de}}}(z) + 3H(1+w_{\mathrm{de}}(z))\rho_{\mathrm{de}}(z)=0.
	\label{Eq:DEcon}
\end{align}
The overdot notation here represents the derivative with respect to cosmic time. From the equation above, one can derive a following relation between PEDE's EoS and density:
\begin{align}
	w_{\mathrm{de}}(z) = -1 + \frac{1}{1+z} \times \frac{\mathrm{d} \ln{\Omega_{\mathrm{de}}(z)}}{\mathrm{d}z},
\end{align}
substituting Eq.~(\ref{Eq:Omde}) in it, we obtain an explicitly PEDE EoS as follows~\cite{li2019simple,pan2020reconciling}
\begin{align}
	w_{\mathrm{de}}(z) = -1 - \frac{1}{3\ln10} \times [1+\tanh(\log_{10}(1+z))].
\end{align}
The equation above reveals that the PEDE equation of state (EoS) exhibits an intriguing symmetry. Specifically, in the distant past, when $z \rightarrow \infty$, the value of $w_{\mathrm{de}}$ approaches $-1 - \frac{2}{3\ln10}$. In the far future, as $z \rightarrow -1$, $w_{\mathrm{de}}$ tends toward $-1$. At the present time, when $z = 0$, we observe that $w_{\mathrm{de}} = -1 - \frac{1}{3\ln10}$, indicating a phantom-like dark energy (DE) EoS. As mentioned briefly in Ref.~\citep{li2019simple}, the pivot redshift at which the matter and dark energy densities are equal marks the transition point for the PEDE EoS.

In the conformal Newtonian gauge, the perturbed FRW metric is given by: \begin{align} ds^2 = a^2(\tau)[-(1+2\psi)d\tau^2 + (1-2\phi)d\vec{r}^2], \end{align} where $\psi$ and $\phi$ denote the metric potentials, and $\vec{r}$ represents the three spatial coordinates. Considering the first-order perturbed components of the conserved stress-energy tensor, the following continuity and Euler equations  for GDM and PEDE can be derived~\citep{kumar2019testing}.
\begin{eqnarray}
	\label{perturbation}
	\delta_{\rm ds}^{\prime}&=& - (1+w_{\rm ds}) \left(\theta_{\rm ds}-3\phi^{\prime}\right)- 3 \mathcal{H} \left(\frac{\delta p_{\rm ds}}{\delta\rho_{\rm ds}} - w_{\rm ds}
	\right)\delta_{\rm ds}\hspace{0.5cm}\\
	\theta_{\rm ds}^{\prime} &=&-\mathcal{H}
	(1-3c_\mathrm{\rm ad,ds}^2)\theta_{\rm ds}  + \frac{\delta p_{\rm ds}/\delta\rho_{\rm ds}}{1+w_{\rm ds}}k^2\delta_{\rm ds} + k^2\psi
\end{eqnarray}
In this context, a prime denotes the derivative with respect to conformal time, $\mathcal{H}$ represents the conformal Hubble parameter, and $k$ refers to the magnitude of the wavevector $\vec{k}$. $\delta_{\rm ds}$ and $\theta_{\rm ds}$ represent the perturbations in the relative density and velocity divergence of the dark sector (DS:DM or DE), $w_{\rm ds}$ and $c_{\rm ad,ds}^2=\frac{p_{\rm ds}^{\prime}}{\rho_{ds}^{\prime}}=w_{\rm ds}-\frac{w_{\rm ds}^{\prime}}{3\mathcal{H}\left(1+w_{\rm ds}\right)}$ denote the DS EoS and the square of DS adiabatic sound speed, respectively. And $\frac{\delta p_{\rm ds}}{\delta\rho_{\rm ds}}$ is the DS sound speed in the Newtonian gauge, it can be expressed as:
\begin{equation}
	\label{deltaP}
	\frac{\delta p_{\rm ds}}{\delta\rho_{\rm ds}}=c_\mathrm{\rm s,ds}^2+3\mathcal{H}\left(1+w_{\rm ds}\right)\left(c_\mathrm{\rm s,ds}^2-c_\mathrm{\rm ad,ds}^2\right)\frac{\theta_{\rm ds}}{k^2},
\end{equation}
here $c_\mathrm{\rm s,ds}^2=c_\mathrm{\rm ad,ds}^2+c_\mathrm{\rm nad,ds}^2$ is the square of DS sound speed in the rest frame. And $c_\mathrm{\rm nad,ds}^2$ represents the square of the non-adiabatic sound speed of the dark sector (DS), describing its micro-scale properties, which must be specified separately. In this work, we consider $c_\mathrm{\rm nad,dm}^2=0$
(Therefore $c_\mathrm{\rm s,dm}^2=c_\mathrm{\rm ad,dm}^2=w_{\rm dm}$) and $c_\mathrm{\rm s,de}^2=1$, then the continuity and Euler equations for GDM and PEDE can be rewritten as:
\begin{eqnarray}
	\label{perturbation}
	\delta_{\rm dm}^{\prime}&=& - (1+w_{\rm dm}) \left(\theta_{\rm dm}-3\phi^{\prime}\right)\\
	\theta_{\rm dm}^{\prime} &=&-\mathcal{H}(1-3w_{\rm dm})\theta_{\rm dm}  + \frac{w_{\rm dm}}{1+w_{\rm dm}}k^2\delta_{\rm dm} + k^2\psi\hspace{0.5cm}
\end{eqnarray}
\begin{align}
	\dot{\delta}_{\mathrm{de}}= & -(1+w_{\mathrm{de}}) \left(\theta_{\mathrm{de}} - 3 \dot{\phi} \right) -3 \mathcal{H}(1 - w_{\mathrm{de}})\delta_{\mathrm{de}}  -3\mathcal{H}w_{\mathrm{de}}^{\prime}
	\frac{\theta_{\mathrm{de}}}{k^2}\notag                           \\
	&- 9 (1+w_{\mathrm{de}})(1- w_{\mathrm{de}})\mathcal{H}^2 \frac{\theta_{\mathrm{de}}}{k^2}, \\
	\dot{\theta}_{\mathrm{de}}= & 2\mathcal{H} \theta_{\mathrm{de}}  + \frac{1}{1+w_{\mathrm{de}}}k^2 \delta_{\mathrm{de}} + k^2\psi.
\end{align}
We still need to provide the initial conditions for DM. In practical calculations, we first give the initial conditions for DM in the synchronous gauge, and then use the transformation rules between the Newtonian gauge and the synchronous gauge to derive the initial conditions in the Newtonian gauge. Since the dark matter is not cold at present, we cannot eliminate the redundant gauge degrees of freedom in the synchronous gauge by setting the velocity divergence of dark matter to zero. However, we can assume that there exists a density-infinitesimal CDM in the universe and set its velocity divergence to zero. Then, we can provide the initial conditions under this gauge, which are as follows:
\begin{align}
	\delta_{\rm dm} =& -\frac{C}{2}(1+w_{\rm dm})(k\tau)^{2},\\
	\theta_{\rm dm}= & -\frac{C}{2}\frac{w_{\rm dm}}{4-3w_{\rm dm}}(k\tau)^{3}k,
\end{align}
where C is a constant. With the equations and initial conditions outlined above, the fundamental background and perturbation dynamics of our new model (denoted as PEDE+$w$DM) are now well understood. We also treat free-streaming neutrinos as a combination of two massless species and one massive species with a mass of $M_v=0.06$~eV. Thus, the complete set of baseline parameters for the PEDE+$w$DM model is provided by:
\begin{equation}\label{}
	\mathcal{P}=\{\omega_b, \omega_{\rm dm}, \theta_s, A_s, n_s, \tau_{\rm reio}, w_{\rm dm}\}.
\end{equation}

\begin{figure*}
	\begin{minipage}{0.5\linewidth}
		\centerline{\includegraphics[width=1\textwidth]{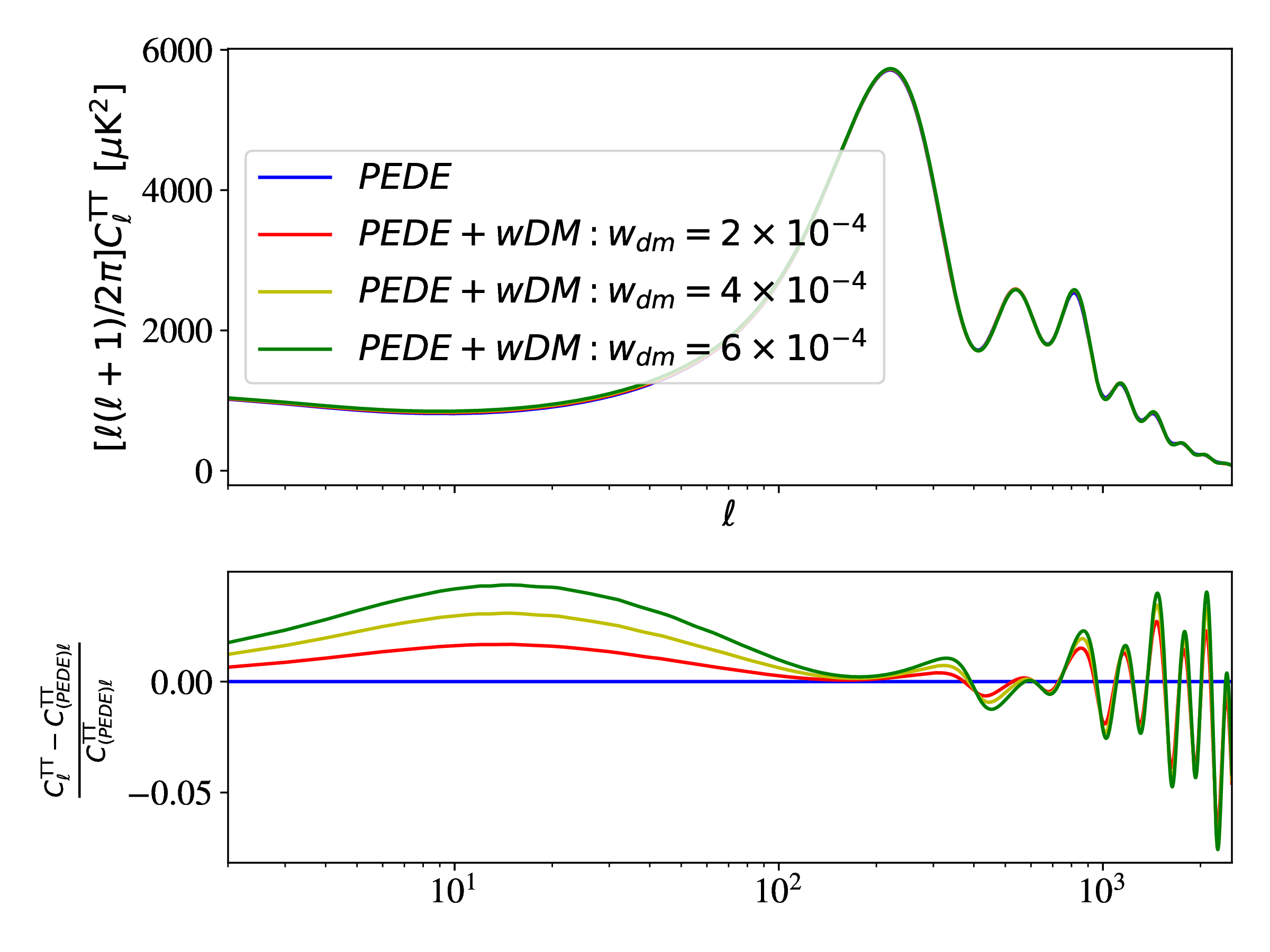}}
		\label{a}
	\end{minipage}
	\begin{minipage}{0.5\linewidth}
		\centerline{\includegraphics[width=1\textwidth]{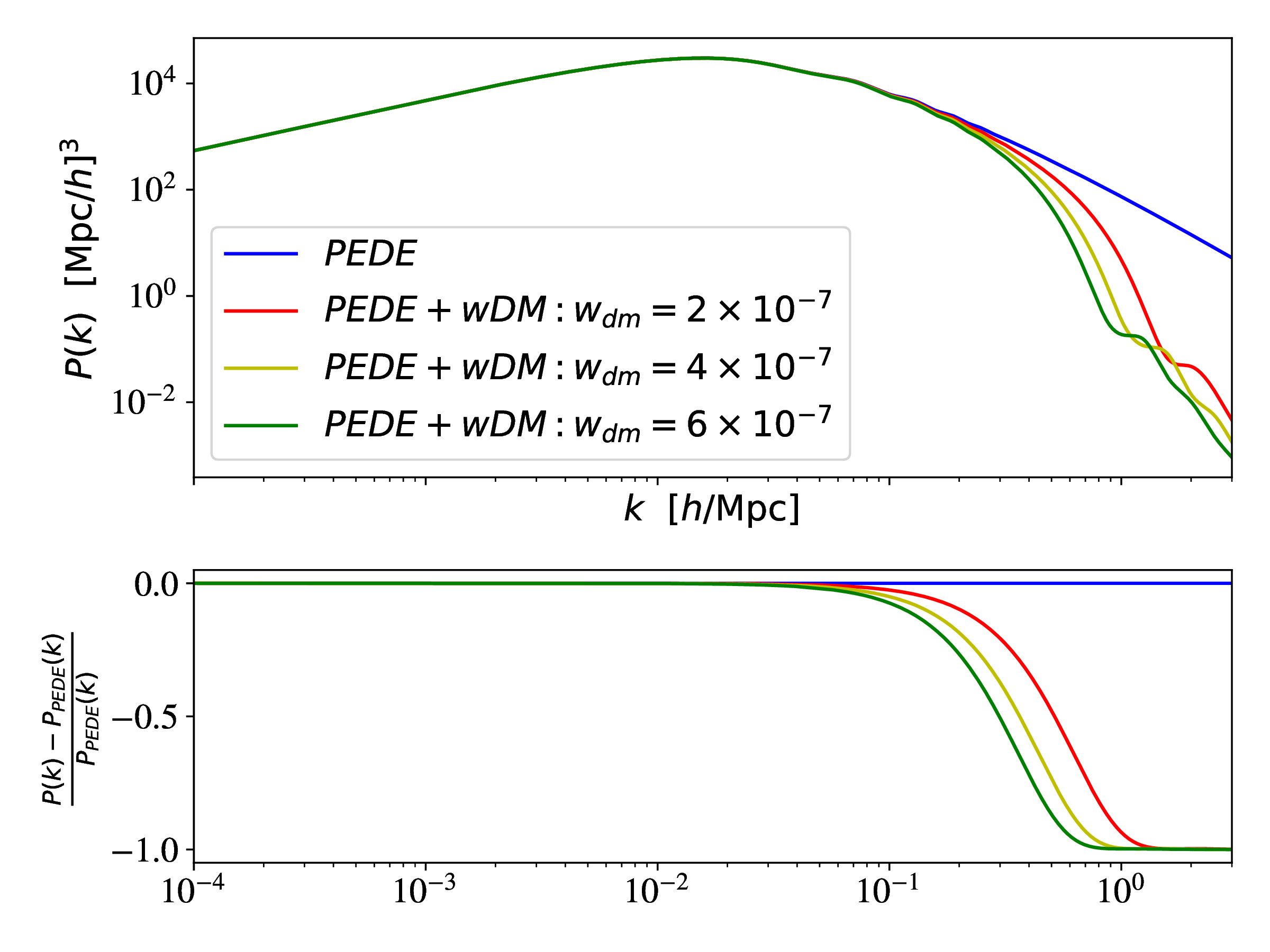}}
	\end{minipage}
	\caption{The CMB TT and matter power spectra for different values of parameter $w_{dm}$, where other relevant model parameters are fixed to their mean values extracted from CMB+DESI+PP+RSD data analysis.}
	\label{fig:1}
\end{figure*}

At the conclusion of this section, we provide an analysis of the effects of the PEDE+$w$DM model on the CMB TT and matter power spectra across various values of wdm. Fig.~\ref{fig:1} illustrates the CMB TT by setting $w_{\rm dm}$ = 0, $2\times 10^{-4}$, $4\times 10^{-4}$, $6\times 10^{-4}$ and matter power spectra by setting $w_{\rm dm}$ = 0, $2\times 10^{-7}$, $4\times 10^{-7}$, $6\times 10^{-7}$ while keeping six other parameters fixed at their mean values derived from CMB+DESI+PP+RSD data analysis.

In the context of CMB TT power spectrum, we observe that the positive values of $w_{\rm dm}$ in the PEDE+$w$DM model result in several percent of relative errors in the CMB TT power spectrum (relative to the case where $w_{\rm dm}=0$) that oscillate when $l>100$, and these relative error oscillations become larger as $l$ increases. At large scales, where the integrated Sachs-Wolfe effect predominates, the primary impact of $w_{\rm dm} > 0$ is an enhanced decay of the gravitational potential from the recombination era to the present, leading to increased anisotropy for $l<100$. For the matter power spectrum, we can see that positive values of $w_{\rm dm}$ decrease the matter power spectrum.

\section{datasets and methodology}\label{data}

To constrain the free parameters of the PEDE+$w$DM model, we rely on the observational datasets presented below.

\textbf{Cosmic Microwave Background (CMB)}: in this research, we employ the Planck 2018 CMB data~\citep{aghanim2020planck-1,aghanim2020planck-2}, focusing on the CMB temperature and polarization angular power spectra plikTTTEEE+low$l$+lowE. Furthermore, we integrate the Planck 2018 CMB lensing reconstruction likelihood~\citep{aghanim2020planck-3} into our analysis.

\textbf{Dark Energy Spectroscopic Instrument (DESI)}: the BAO measurements from the first year of DESI observations are incorporated, as described in Ref.\cite{adame2024desi1}. This dataset includes the Bright Galaxy Sample (BGS), LRG, a combination of LRG and ELG, ELG, QSO, and Ly$\alpha$ forest\cite{adame2024desi1,adame2024desi2,adame2024desi3}.

\textbf{Pantheon Plus (PP)}: the distance modulus measurements of SNe Ia from the PP sample, detailed in~\citep{scolnic2022pantheon+}, are used in this work. This dataset consists of 1701 light curves from 1550 distinct SNe Ia events, spanning redshifts between $z\in[0.001, 2.26]$.

\textbf{Redshift Space Distortions (RSD)}: RSD is a crucial tool for probing the combination $f\sigma_8$. In the PEDE+$w$DM model, the quantity $f$ is scale-dependent and is defined as:
\begin{equation}\label{}
	f(k,a)=\frac{d\ln\delta(k,a)}{d\ln a}, \hspace{1cm} \delta(k,a)=\sqrt{\frac{P(k,a)}{P(k,a_0)}},
\end{equation}
where $P(k,a)$ denotes the matter power spectrum, given by:
\begin{equation}\label{}
	\langle\widetilde{\delta}_m(\textbf{k},t)\widetilde{\delta}_m^\ast(\textbf{k}^{\prime},t)\rangle=(2\pi)^3P(\textbf{k},a(t))\delta^3(\textbf{k}-\textbf{k}^{\prime}),
\end{equation}
with $\widetilde{\delta}m(\textbf{k},t)$ being the Fourier transform of $\delta_m=\frac{\delta\rho{\rm dm}+\delta\rho_b+\delta\rho_{\rm \overline{\nu}}}{\bar{\rho}_{\rm dm}+\bar{\rho}b+\bar{\rho}{\rm \overline{\nu}}}$, where $\overline{\nu}$ refers to massive neutrinos. The value of $\sigma_8$ is calculated as:
\begin{equation}\label{}
	\sigma_8(a)=\sqrt{\int_0^{+\infty}dk\frac{k^2P(k,a)W_R^2(k)}{2\pi^2}}
\end{equation}
where $W_R(k) = 3[\sin(kR)/kR-\cos(kR)]/(kR)^2$ is the Fourier transform of the top-hat window function, and $R$ determines the scale for computing the root-mean-square (RMS) normalization of matter fluctuations.

This work uses the RSD measurements of $f\sigma_8$ from Table I of Ref.~\cite{sagredo2018internal}. This dataset, referred to as the "Gold 2018" sample in the literature, includes 22 measurements of $f\sigma_8$ within the redshift range $0.02<z<1.944$~\cite{song2009reconstructing,davis2011local,samushia2012interpreting,turnbull2012cosmic,hudson2012growth,blake2012wigglez,blake2013galaxy,sanchez2014clustering,chuang2016clustering,howlett2015clustering,feix2015growth,okumura2016subaru,huterer2017testing,pezzotta2017vimos,zhao2019clustering}. For our analysis, we use the publicly available Gold 2018 Montepython likelihood~\footnote{https://github.com/snesseris/RSD-growth} for the $f\sigma_8$ measurements. In this likelihood, the wavenumber $k$ in the definition of $f$ is fixed at 0.1 Mpc, aligning with the effective wavenumber of the RSD measurements used.

To constrain the PEDE+$w$DM model, we employ a Markov Chain Monte Carlo (MCMC) method, utilizing the public MontePython-v3 code~\cite{audren2013conservative,brinckmann2019montepython} and a modified version of the CLASS code~\cite{lesgourgues2011cosmic,blas2011cosmic}. The analysis is carried out using the Metropolis-Hastings algorithm, and we assess convergence of the chains using the Gelman-Rubin criterion~\cite{gelman1992inference}, requiring $R-1<0.01$. The flat priors on the free parameters of the PEDE+$w$DM model are provided in Table~\ref{tab:prior}. In particular, since the condition $c_{\rm s,dm}^2=w_{\rm dm}<0$ will lead to an imaginary value of sound speed of DM, we impose a lower bound of 0 for the flat prior on the parameter $w_{\rm dm}$.
\begin{table}
\begin{center}
\caption[]{Flat priors for the free parameters of the PEDE+$w$DM model.}
\label{tab:prior}
\begin{tabular}{c c}
	\hline
				Parameters               & Prior       \\
				\hline
				$100\omega{}_{b}$        & [0.8,2.4]   \\
				$\omega{}_{\mathrm{dm}}$ & [0.01,0.99] \\
				$100\theta{}_{s}$        & [0.5,2.0]   \\
				$\ln[10^{10}A_{s}]$      & [2.7,4.0]   \\
				$n_{s}$                  & [0.9,1.1]   \\
				$\tau{}_{\mathrm{reio}}$ & [0.01,0.8]  \\
				$10^{6}w_{\mathrm{dm}}$  & [0,100]     \\ \hline

\end{tabular}
\end{center}
\end{table}

Finally, we will analyze the performance of the PEDE+$w$DM model in comparison to the $\Lambda$CDM model and the PEDE model. We utilized the cosmological code MCEvidence, developed by the authors of Ref.~\citep{heavens2017marginal, heavens2017no}, to calculate Bayesian evidence for all datasets and referred to Ref.~\citep{pan2018observational, yang2019constraints} for further discussion on this topic. The performance of a cosmological model (denoted as $M_{i}$) with respect to a reference cosmological model is quantified by the Bayes factor $B_{ij}$ (or its logarithm $B_{ij}$) of the model $M_i$ with respect to the reference model $M_j$. In Table~\ref{tab:Bij}, we present the revised Jeffrey's scale~\cite{kass1995bayes} which quantifies the strength of evidence for model $M_i$ compared to model $M_j$.

\begin{table}
\begin{center}
	\caption[]{Revised Jeffreys' scale, used to measure the relative strength of evidence for model $M_i$ over model $M_j$.}
\label{tab:Bij}
			\begin{tabular}{c c} \hline
				$\ln B_{ij}$            & Strength of evidence for model $M_i$ \\
				\hline
				$0 \leq \ln B_{ij} < 1$ & Weak                                 \\
				$1 \leq \ln B_{ij} < 3$ & Definite/Positive                    \\
				$3 \leq \ln B_{ij} < 5$ & Strong                               \\
				$\ln B_{ij} \geq 5$     & Very strong                          \\ \hline
			\end{tabular}
\end{center}
\end{table}

\section{results and discussion}\label{results}

\begin{figure}
	\centering
	\includegraphics[width=0.8\textwidth,angle=0]{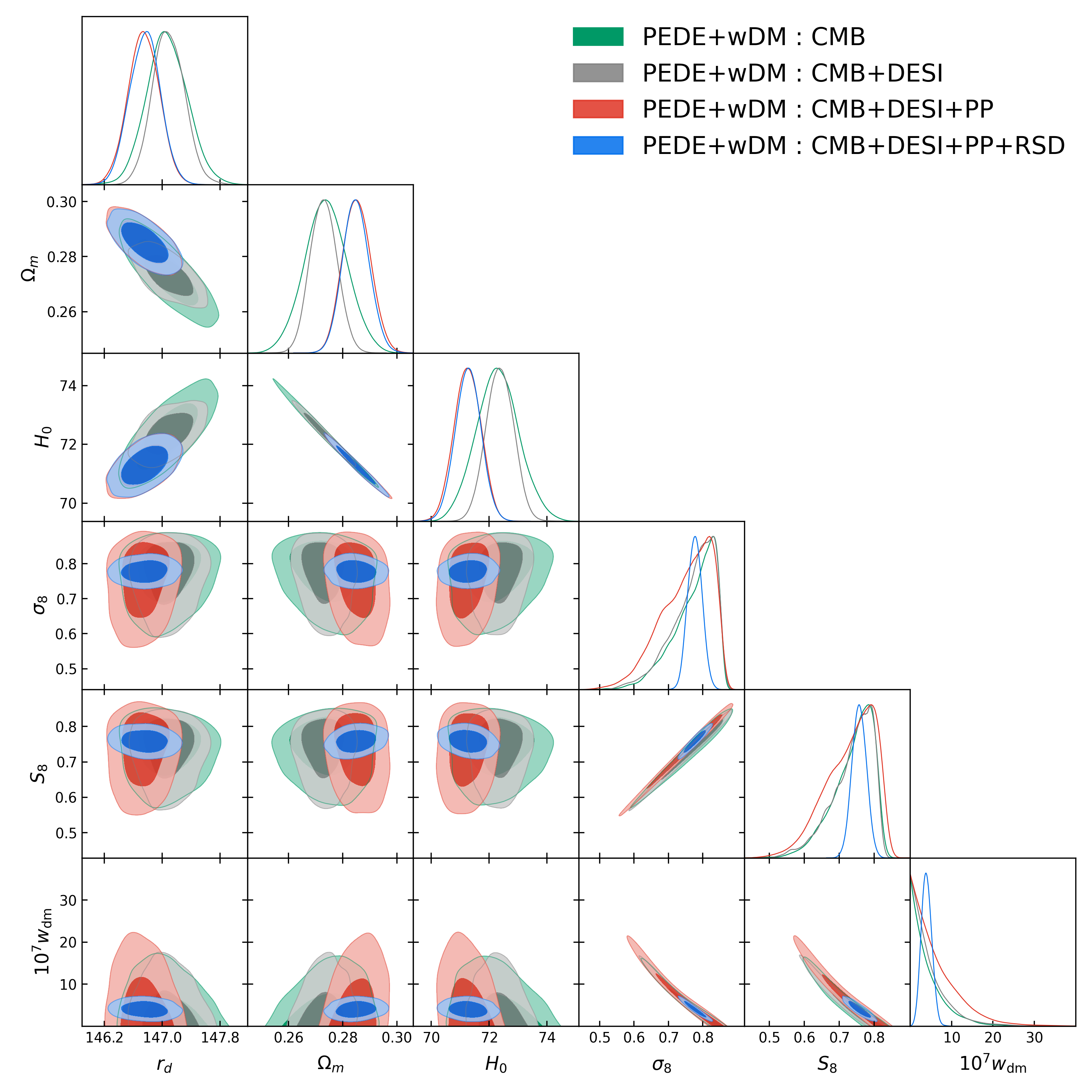}
	\caption{The posterior distributions (one-dimensional) and the two-dimensional joint confidence contours (68\% and 95\%) for the most important parameters of the PEDE+$w$DM model are shown using the CMB, CMB+DESI, CMB+DESI+PP, and CMB+DESI+PP+RSD observational datasets. }
	\label{fig:PEDE+wDM}
\end{figure}

\begin{figure}
	\centering
	\includegraphics[width=0.8\textwidth,angle=0]{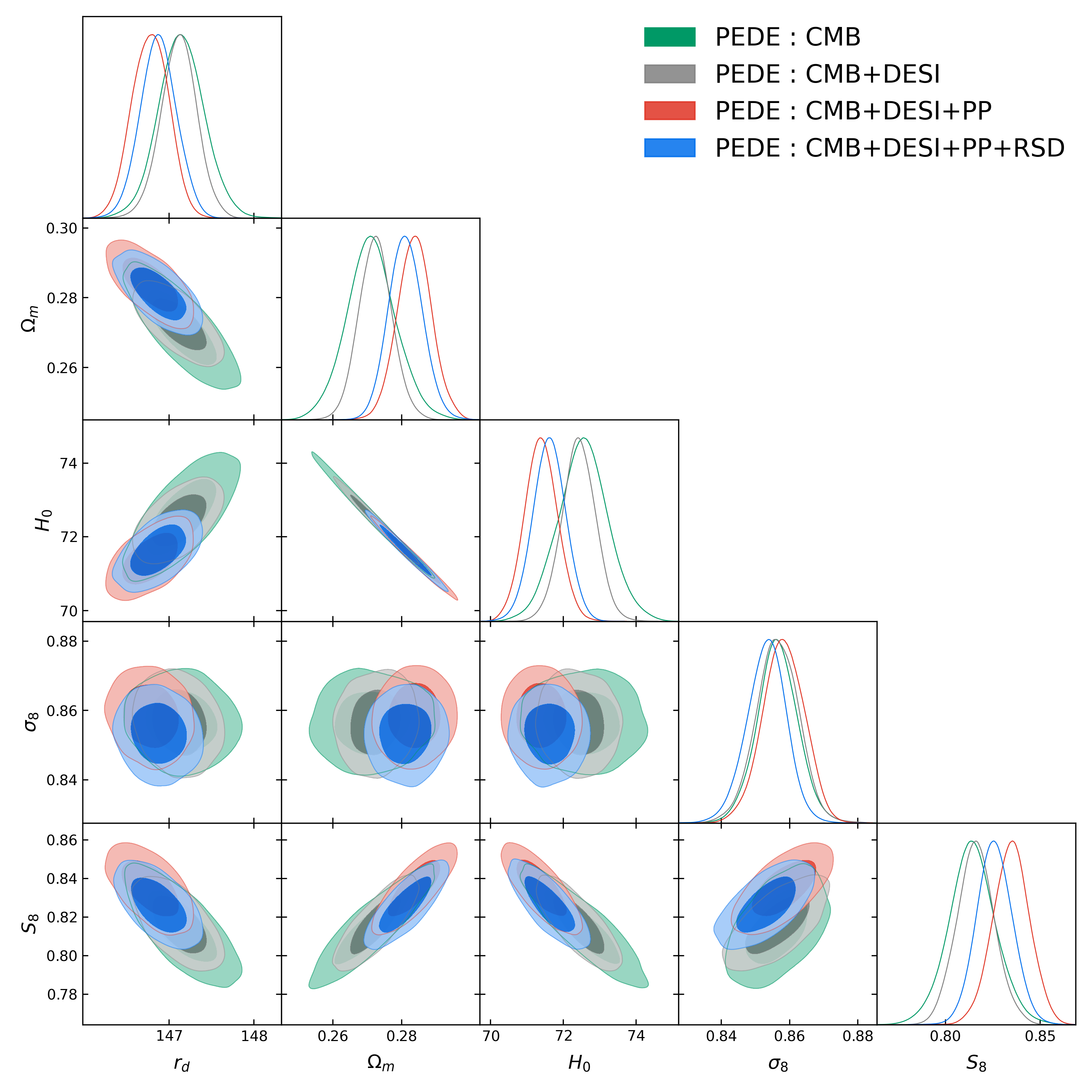}
	\caption{The posterior distributions (one-dimensional) and the two-dimensional joint confidence contours (68\% and 95\%) for the most important parameters of the PEDE model are shown using the CMB, CMB+DESI, CMB+DESI+PP, and CMB+DESI+PP+RSD observational datasets.}
	\label{fig:PEDE}
\end{figure}

The constraints on the PEDE+$w$DM model and the PEDE model for the CMB, CMB+DESI, CMB+DESI+PP, and CMB+DESI+PP+RSD datasets are presented in Tables~\ref{tab:PEDE+wDM},\ref{tab:PEDE} and Figures\ref{fig:PEDE+wDM},~\ref{fig:PEDE}.

We start by examining the fitting results of the PEDE+$w$DM model using only CMB data. Subsequently, we explore the impact of incorporating additional probes by progressively adding them to the CMB analysis. Using only CMB data, we find no statistically significant signal for a positive DM parameter $w_{\rm dm}$ ($0<10^{7}w_{\rm dm}<13.1$ at the 95\% confidence level). Nevertheless, differences in parameters $\sigma_8$ and $S_8$  between the PEDE model and the PEDE+$w$DM model are presented due to the small but not negligible positive mean value of $w_{\rm dm}$. More specifically, the small but non-vanish positive mean value of parameter $w_{\rm dm}$ reduces the parameters $\sigma_8$ and $S_8$ from $\sigma_8=0.8565\pm0.0060$ (at the 68\% confidence level) and $S_8=0.814\pm0.013$ (at the 68\% confidence level) in the PEDE model to $\sigma_8=0.776^{+0.077}_{-0.029}$ (at the 68\% confidence level) and $S_8=0.741^{+0.072}_{-0.031}$ (at the 68\% confidence level) in the PEDE+$w$DM model. We point out that these changes mainly originate from the effects of a non-zero mean value of DM sound speed squared rather than directly from the effects of a DM EoS parameter, because the value of DM EoS parameter on the order of $10^{-7}$ has a negligible direct effect on the fitting results of other parameters. One notes that the posteriors of parameters $\sigma_8$ and $S_8$ exhibit non-Gaussianity, this is because the parameter $w_{\rm dm}$ is restricted to be positive. A positive value of $w_{\rm dm}$ is necessary since otherwise DM will encounter instability.  For the other parameters, introducing a DM parameter $w_{\rm dm}$ does not cause them to undergo significant changes. For example, the introduction of $w_{\rm dm}$ only reduce $H_0$ from $H_0=72.5\pm0.70$ (at the 68\% confidence level) in the PEDE model to $H_0=72.33\pm 0.75$ (at the 68\% confidence level) in the PEDE+$w$DM model.

With the addition of DESI data to CMB, we observe little change in the fitting results. In particular, we still find no statistically significant signal for a positive DM parameter $w_{\rm dm}$ ($0<10^{7}w_{\rm dm}<13.2$ at the 95\% confidence level). Furthermore, the small but not negligible positive mean value of $w_{\rm dm}$ leads to similar changes on parameters $\sigma_8$, $S_8$ and $H_0$ between PEDE+$w$DM and PEDE with respect to the fitting results regarding CMB dataset alone.

When PP are added to CMB+DESI, we still find no statistically significant signal for a positive DM parameter $w_{\rm dm}$  ($0<10^{7}w_{\rm dm}<17.6$ at the 95\% confidence level). In addition, we find that, compared to the fitting results regarding CMB+DESI dataset, the mean vaules of parameters $H_0$ and $\Omega_m$ change from $H_0\sim72$ and $\Omega_m\sim0.27$ to $H_0\sim71$ and $\Omega_m\sim0.28$.

For the CMB+DESI+PP+RSD dataset, we finally find statistically significant signal for a positive DM parameter $w_{\rm dm}$ ($10^{7}w_{\rm dm}$ = $4.0^{+2.5}_{-2.3}$ at the 95\% confidence level). Interestingly, this result is seemingly opposite to the result in our previous article~\citep{yao2024observational}, in which the CMB+Pantheon and CMB+BAO+Pantheon datasets gave a result that the DM EoS is less than 0 at the 95\% confidence level. There are two reason for the seemingly opposite results in the two articles: First, in the previous article, we did not use the RSD data, and the BAO and SN Ia data were also outdated. From the posterior distribution in Fig.~\ref{fig:PEDE+wDM}, it can be seen that even in this article, when the RSD dataset is not included, the extremum of $w_{\rm dm}$ is also less than 0. Second, the models in the two articles are different. In the previous article, we set the sound speed of DM to 0, so even if we had included the RSD data at that time, it would not have been likely to result in a positive $w_{\rm dm}$. However, in this article, we set the square of the DM sound speed equal to its EoS parameter. Since the RSD data supports a smaller $S_8$ parameter than that predicted by the CMB, and the DM sound speed is negatively correlated with the $S_8$ parameter, the RSD data supports a slightly positive DM sound speed, which corresponds to a slightly positive DM EoS parameter in this model. Additionally, apart from the mean values of the parameters $\sigma_8$ and $S_8$ being increasd by a lower mean value of the parameter $w_{\rm dm}$ imposed by the RSD dataset, the other model parameters remain largely unchanged.

In the final step, we provide the $\ln B_{ij}$ values, which quantify the evidence for the fit of the PEDE model and the PEDE+$w$DM model relative to the $\Lambda$CDM model, based on the datasets considered in this work, in Table~\ref{tab:lnB_LCDM}. Referring to the Revised Jeffreys' scale presented in Table~\ref{tab:Bij}, the Bayesian evidence indicates that all datasets analyzed here favor the $\Lambda$CDM model over the PEDE+$w$DM model. More specifically, the CMB and CMB+DESI datasets favor $\Lambda$CDM over PEDE+$w$DM with strong evidence, and the CMB+DESI+PP and CMB+DESI+PP+RSD datasets favor $\Lambda$CDM over PEDE+$w$DM with very strong evidence. In addition, while CMB+DESI+PP+RSD dataset favor PEDE+$w$DM over PEDE with weak evidence, other datasets favor PEDE over PEDE+$w$DM with strong or very strong evidence.

\begin{table}
\begin{center}
	\caption[]{The mean values along with the 1$\sigma$ and 2$\sigma$ ranges for the PEDE+$w$DM model based on the CMB, CMB+DESI, CMB+DESI+PP, and CMB+DESI+PP+RSD datasets.}
\label{tab:PEDE+wDM}
\scalebox{0.8}{			\begin{tabular}{ccccc}\hline
				Parameters                         & CMB                                              & CMB+DESI                                     & CMB+DESI+PP                                  & CMB+DESI+PP+RSD                              \\
				\hline
				{\boldmath$100\omega{}_{b }$}      & ${2.235\pm 0.015}^{+0.031}_{-0.029}$             & ${2.236\pm 0.013}^{+0.027}_{-0.027}$         & ${2.220\pm 0.013}^{+0.025}_{-0.026}$         & ${2.220\pm 0.013}^{+0.027}_{-0.025}$         \\
				{\boldmath$\omega{}_{\rm cdm }$}   & ${0.1201\pm 0.0013}^{+0.0026}_{-0.0026}$         & ${0.12003\pm 0.00085}^{+0.0016}_{-0.0016}$   & ${0.12202\pm 0.00085}^{+0.0017}_{-0.0016}$   & ${0.12195\pm 0.00081}^{+0.0016}_{-0.0016}$   \\
				{\boldmath$100\theta{}_{s }$}      & ${1.04187\pm 0.00030}^{+0.00060}_{-0.00058}$     & ${1.04191\pm 0.00028}^{+0.00055}_{-0.00055}$ & ${1.04177\pm 0.00028}^{+0.00054}_{-0.00053}$ & ${1.04175\pm 0.00028}^{+0.00055}_{-0.00054}$ \\
				{\boldmath$\ln(10^{10}A_{s })$}    & ${3.045^{+0.014}_{-0.015}}^{+0.029}_{-0.028}$    & ${3.047\pm 0.015}^{+0.031}_{-0.029}$         & ${3.041\pm 0.014}^{+0.027}_{-0.027}$         & ${3.039\pm 0.014}^{+0.027}_{-0.027}$         \\
				{\boldmath$n_{s }$}                & ${0.9647\pm 0.0044}^{+0.0086}_{-0.0085}$         & ${0.9648\pm 0.0036}^{+0.0072}_{-0.0070}$     & ${0.9599\pm 0.0036}^{+0.0072}_{-0.0068}$     & ${0.9602\pm 0.0035}^{+0.0067}_{-0.0068}$     \\
				{\boldmath$\tau{}_{\rm reio }$}    & ${0.0547^{+0.0071}_{-0.0079}}^{+0.015}_{-0.015}$ & ${0.0554\pm 0.0076}^{+0.015}_{-0.015}$       & ${0.0506\pm 0.0070}^{+0.014}_{-0.013}$       & ${0.0499\pm 0.0068}^{+0.013}_{-0.013}$       \\
				{\boldmath$10^{7}w_{\mathrm{dm}}$} & $0< 5.07< 13.1$                                   & $0< 5.17< 13.2 $                              & $0< 7.54< 17.6 $                              & ${4.0^{+1.2}_{-1.3}}^{+2.5}_{-2.3}$          \\
				\hline
				               {\boldmath$r_{d }$}                & ${147.07\pm 0.28}^{+0.57}_{-0.54}$               & ${147.09\pm 0.22}^{+0.43}_{-0.43}$           & ${146.75\pm 0.22}^{+0.42}_{-0.43}$           & ${146.76\pm 0.21}^{+0.42}_{-0.41}$           \\
				{\boldmath$H_0$}                   & ${72.31\pm 0.74}^{+1.5}_{-1.4}$                  & ${72.38\pm 0.47}^{+0.90}_{-0.92}$            & ${71.26\pm 0.46}^{+0.88}_{-0.90}$            & ${71.28\pm 0.44}^{+0.87}_{-0.86}$            \\
				{\boldmath$\Omega{}_{m }$}         & ${0.2739\pm 0.0078}^{+0.016}_{-0.015}$           & ${0.2731\pm 0.0050}^{+0.0099}_{-0.0094}$     & ${0.2853\pm 0.0051}^{+0.010}_{-0.0095}$      & ${0.2850\pm 0.0049}^{+0.0099}_{-0.0092}$     \\
				{\boldmath$\sigma_8$}              & ${0.776^{+0.077}_{-0.029}}^{+0.088}_{-0.13}$     & ${0.774^{+0.079}_{-0.030}}^{+0.093}_{-0.13}$ & ${0.753^{+0.096}_{-0.043}}^{+0.11}_{-0.14}$  & ${0.778\pm 0.021}^{+0.041}_{-0.039}$         \\
				{\boldmath$S_8$}                   & ${0.741^{+0.072}_{-0.031}}^{+0.089}_{-0.12}$     & ${0.738^{+0.075}_{-0.027}}^{+0.088}_{-0.12}$ & ${0.734^{+0.091}_{-0.043}}^{+0.11}_{-0.14}$  & ${0.758\pm 0.021}^{+0.040}_{-0.040}$         \\
				\hline
				{\boldmath$\chi^2_{\mathrm{min}}$} & 2780.26                                          & 2796.82                                      & 4251.68                                      & 4265.76                                      \\ \hline
			\end{tabular}}
	\end{center}
\end{table}

\begin{table}
\begin{center}
\caption[]{The mean values along with the 1$\sigma$ and 2$\sigma$ ranges for the PEDE model based on the CMB, CMB+DESI, CMB+DESI+PP, and CMB+DESI+PP+RSD datasets.}
\label{tab:PEDE}
\scalebox{0.8}{			\begin{tabular}{ccccc}\hline
				Parameters                         & CMB                                          & CMB+DESI                                     & CMB+DESI+PP                                      & CMB+DESI+PP+RSD                                  \\
				\hline
				{\boldmath$100\omega{}_{b }$}      & ${2.239\pm 0.015}^{+0.029}_{-0.029}$         & ${2.237\pm 0.013}^{+0.026}_{-0.026}$         & ${2.222\pm 0.013}^{+0.024}_{-0.024}$             & ${2.224\pm 0.013}^{+0.024}_{-0.026}$             \\
				{\boldmath$\omega{}_{\rm cdm }$}   & ${0.1197\pm 0.0012}^{+0.0024}_{-0.0024}$     & ${0.11990\pm 0.00084}^{+0.0017}_{-0.0017}$   & ${0.12178\pm 0.00084}^{+0.0017}_{-0.0017}$       & ${0.12133\pm 0.00083}^{+0.0016}_{-0.0016}$       \\
				{\boldmath$100\theta{}_{s }$}      & ${1.04190\pm 0.00031}^{+0.00058}_{-0.00062}$ & ${1.04188\pm 0.00027}^{+0.00052}_{-0.00056}$ & ${1.04177\pm 0.00028}^{+0.00056}_{-0.00054}$     & ${1.04177\pm 0.00028}^{+0.00055}_{-0.00056}$     \\
				{\boldmath$\ln(10^{10}A_{s })$}    & ${3.044\pm 0.015}^{+0.030}_{-0.029}$         & ${3.043\pm 0.014}^{+0.027}_{-0.029}$         & ${3.037\pm 0.014}^{+0.026}_{-0.027}$             & ${3.028^{+0.014}_{-0.012}}^{+0.025}_{-0.028}$    \\
				{\boldmath$n_{s }$}                & ${0.9657\pm 0.0041}^{+0.0088}_{-0.0077}$     & ${0.9654\pm 0.0037}^{+0.0072}_{-0.0070}$     & ${0.9607\pm 0.0035}^{+0.0069}_{-0.0066}$         & ${0.9615\pm 0.0035}^{+0.0066}_{-0.0069}$         \\
				{\boldmath$\tau{}_{reio }$}        & ${0.0546\pm 0.0076}^{+0.016}_{-0.015}$       & ${0.0537\pm 0.0072}^{+0.014}_{-0.014}$       & ${0.0493^{+0.0073}_{-0.0065}}^{+0.013}_{-0.014}$ & ${0.0455^{+0.0072}_{-0.0063}}^{+0.013}_{-0.014}$ \\
				\hline
				               {\boldmath$r_{d }$}                & ${147.15\pm 0.27}^{+0.55}_{-0.53}$           & ${147.12\pm 0.21}^{+0.43}_{-0.43}$           & ${146.78\pm 0.22}^{+0.41}_{-0.42}$               & ${146.88\pm 0.21}^{+0.42}_{-0.42}$               \\
				{\boldmath$H_0$}                   & ${72.55\pm 0.69}^{+1.4}_{-1.3}$              & ${72.43\pm 0.47}^{+0.91}_{-0.94}$            & ${71.40\pm 0.45}^{+0.92}_{-0.89}$                & ${71.62\pm 0.45}^{+0.88}_{-0.89}$                \\
				{\boldmath$\Omega{}_{m }$}         & ${0.2713\pm 0.0072}^{+0.014}_{-0.014}$       & ${0.2724\pm 0.0049}^{+0.0098}_{-0.0094}$     & ${0.2838\pm 0.0050}^{+0.010}_{-0.010}$           & ${0.2812\pm 0.0050}^{+0.010}_{-0.0095}$          \\
				{\boldmath$\sigma_8$}              & ${0.8565\pm 0.0060}^{+0.012}_{-0.012}$       & ${0.8565\pm 0.0063}^{+0.012}_{-0.012}$       & ${0.8582\pm 0.0061}^{+0.012}_{-0.012}$           & ${0.8532^{+0.0060}_{-0.0053}}^{+0.011}_{-0.012}$ \\
				{\boldmath$S_8$}                   & ${0.814\pm 0.013}^{+0.026}_{-0.025}$         & ${0.816\pm 0.010}^{+0.021}_{-0.019}$         & ${0.8347\pm 0.0097}^{+0.019}_{-0.020}$           & ${0.8261\pm 0.0094}^{+0.019}_{-0.018}$           \\
				\hline
				{\boldmath$\chi^2_{\mathrm{min}}$} & 2778.78                                      & 2796.5                                       & 4250.70                                          & 4276.80                                          \\ \hline
			\end{tabular}}
	\end{center}
\end{table}

\begin{table}
\begin{center}
		\caption[]{The $\ln B_{ij}$ values summarize the evidence for the PEDE and PEDE+$w$DM models in comparison to the $\Lambda$CDM model, based on the CMB, CMB+DESI, CMB+DESI+PP, and CMB+DESI+PP+RSD datasets. A negative $\ln B_{ij}$ value suggests the PEDE or PEDE+$w$DM models are less favored than the $\Lambda$CDM model.}
	\label{tab:lnB_LCDM}
		\begin{tabular}{c c c} \hline
			Model                  & Datasets        & $\ln B_{ij}$ \\
			\hline
			PEDE                   & CMB             & 0.21         \\
			PEDE                   & CMB+DESI        & 1.56         \\
			PEDE                   & CMB+DESI+PP     & $-$21.13     \\
			PEDE                   & CMB+DESI+PP+RSD & $-$25.65     \\ \hline
			PEDE+$w_{\mathrm{dm}}$ & CMB             & $-$4.99      \\
			PEDE+$w_{\mathrm{dm}}$ & CMB+DESI        & $-$3.68      \\
			PEDE+$w_{\mathrm{dm}}$ & CMB+DESI+PP     & $-$25.93     \\
			PEDE+$w_{\mathrm{dm}}$ & CMB+DESI+PP+RSD & $-$25.14     \\ \hline
		\end{tabular}
	\end{center}
\end{table}

\section{concluding remarks}
\label{conclusion}
Although CDM is strongly supported by a wide range of cosmological observations when paired with a cosmological constant, it faces several persistent challenges on small scales. These include the missing satellites problem, the too-big-to-fail problem, and the core-cusp problem, which suggest potential discrepancies between CDM predictions and observed astrophysical structures.
These unresolved issues have spurred the exploration of alternative dark matter candidates beyond CDM, such as: Warm DM, Fuzzy DM, Interacting DM and Decaying DM. These models aim to suppress the overproduction of low-mass structures in simulations, offering possible solutions to CDM's small-scale anomalies. Nevertheless, the current limitations of CDM leave the fundamental question---"What is dark matter?"---largely unanswered, keeping the field open for further theoretical and observational investigation. In light of this, we have proposed a new cosmological model by considering DM as a barotropic fluid characterized by a constant EoS parameter and DE as PEDE, considering that there have already been many work on the extended properties of DM in the context of a cosmological constant and there are some interesting results emerged as we investigated the extended properties of DM in the context of PEDE.~\citep{yao2024observational}.

Using the background and perturbation equations of the PEDE+$w$DM model, we performed fits to multiple datasets: CMB alone, CMB+DESI, CMB+DESI+PP, and CMB+DESI+PP+RSD. Our analysis reveals: for the CMB, CMB+DESI, and CMB+DESI+PP datasets, we detect no statistically signifiant evidence for a nonzero DM parameter $w_{\rm dm}$. However, the small but non-negligible positive mean value of $w_{\rm dm}$ in these cases results in a modest reduction of the parameters $\sigma_8$ and $S_8$. In contrast, the CMB+DESI+PP+RSD dataset yields a statistically significant detection of a positive $w_{\rm dm}$, indicating a measurable departure from CDM behavior. Finally, Bayesian evidence analysis indicates that, across all datasets examined in this work, $\Lambda$CDM is preferred over PEDE+$w$DM. Specifically, CMB alone and CMB+DESI show strong evidence in favor of $\Lambda$CDM. CMB+DESI+PP and CMB+DESI+PP+RSD exhibit very strong evidence for $\Lambda$CDM over PEDE+$w$DM. Regarding the comparison between PEDE+$w$DM and PEDE, only the CMB+DESI+PP+RSD dataset weakly favors PEDE+$w$DM over PEDE. All other datasets show strong or very strong evidence for PEDE over PEDE+$w$DM.

\begin{acknowledgements}
 This work is supported by the  Guangdong Basic and Applied Basic Research Foundation (Grant No.2024A1515012573), National key R\&D Program of China (Grant No.2020YFC2201600),  National Natural Science Foundation of China (Grant No.12073088), and National SKA Program of China (Grant No. 2020SKA0110402).
\end{acknowledgements}

\bibliographystyle{raa}
\bibliography{ms2025-0126bibtex}

\label{lastpage}

\end{document}